\documentclass[multphys,vecphys]{svmult}

\usepackage{makeidx}     
\usepackage{graphicx}    
\usepackage{multicol}    

\makeindex             

\begin{document}

\title*{Molecular Gas in Nearby Dwarf Galaxies: Single Dish and
Interferometric Results}

\titlerunning{Molecular Gas in Nearby Dwarf Galaxies}


\author{Alberto D. Bolatto\inst{1}, Adam Leroy\inst{1}, Josh
D. Simon\inst{1}, Leo Blitz\inst{1}, \and Fabian Walter\inst{2}}

\authorrunning{Bolatto et al.}


\institute{Department of Astronomy, University of California at Berkeley
\texttt{bolatto@astro.berkeley.edu}
\and National Radio Astronomy Observatory}

%
%
\maketitle


\section{Introduction: The MIDGET Survey}

During the past two years we have undertaken a systematic
millimeter--wave survey of molecular gas in a sample of 150 nearby,
northern dwarf galaxies with IRAS emission.  We have called this
survey MIDGET (Millimeter Interferometry of Dwarf Galaxies).  MIDGET
has two parts: a single--dish search for CO done at the UASO Kitt Peak
12m telescope (Leroy et al., in preparation), and an interferometric
follow-up of the galaxies with strong CO emission carried out at the
Berkeley--Illinois--Maryland Array (BIMA). The single--dish survey
targeted the centers of northern galaxies that are nearby (V$_{\rm
LSR}\leq1000$ km s$^{-1}$), of small mass (H{\small I} linewidths
$W_{20}\leq200$ km s$^{-1}$), compact in size, and detected by
IRAS. The single--dish observations were very successful, finding $41$
new CO emitters and more than doubling the number of previously known
CO sources within the defined sample. The typical $1\sigma$
sensitivity attained was $\sim0.2$ K km s$^{-1}$. Carbon monoxide
emission was detected in galaxies with $12 + \log (\mbox{[O/H]})
\sim7.9-8.5$, a metallicity regime similar to that of the Magellanic
Clouds.  We are using these data to address several questions related
to molecular cloud and star formation in these objects. This paper
concentrates on the issue of how CO traces molecular gas as a function
of heavy element abundance $Z$.

\section{Does the X$_{\rm CO}$ Factor Depend Strongly on Metallicity?}
\label{xcodepz}

\begin{figure}[ht]
\centering \includegraphics[height=5.2cm]{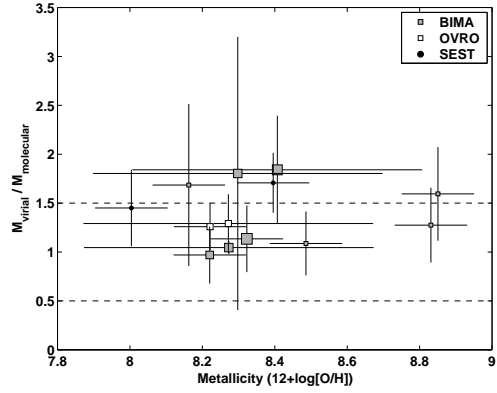}
\caption{Virial CO--to--H$_2$ conversion factor as a function of
metallicity, normalized to the Galactic value. Each point represents
the average of the ratio of virial to molecular mass for all the GMCs
analyzed in one galaxy, with the molecular mass computed using the
integrated CO intensity and the Galactic X$_{\rm CO}$ factor. Symbol
color identifies the telescopes used for the measurements, while size
is proportional to the spatial resolution attained in the
observations.  The data include clouds in M~33, the SMC (N83/N84), the
LMC (N159/N160), IC~10, NGC~2976, NGC~3077, NGC~4214, NGC~4449, and
NGC~4605.}
\label{xco}
\end{figure}

Measuring the amount of molecular hydrogen present using CO
observations requires assuming a CO--to--H$_2$ conversion factor
X$_{\rm CO}=N({\rm H}_2)/\int {\rm I(CO)}\,dv$ (e.g., Sanders,
Solomon, \& Scoville 1984). This factor will be a product of
abundance, excitation, and cloud structure averaged over a large area.
The theoretical expectation is that the X$_{\rm CO}$ factor will
depend on the local properties of the interstellar medium (ISM) such
as volume density, temperature, and metallicity. The dependence of
X$_{\rm CO}$ on metallicity $Z$ remains a point of contention.  Some
authors find X$_{\rm CO}$ strongly increasing for decreasing $Z$
(e.g., Wilson 1995; Israel 1997), while other, recent work finds no
apparent trend (e.g., Rosolowsky et al. 2003).

We have used the interferometric data to measure the properties of
individual molecular clouds (size, velocity dispersion, CO luminosity)
in dwarf galaxies where we can accurately derive virial masses. For
these measurements we have developed a robust algorithm called
{\tt\small CLOUDALYZE}, which uses moments of the intensity
distribution computed over connected regions of the datacube as a
function of a threshold intensity (Bolatto et al. 2003). These
moments, adequately corrected to account for angular resolution and
signal--to--noise effects, are then used to compute the size and
velocity dispersion of a cloud, thus measuring its virial mass.  In
designing this experiment we have explicitly tried to avoid problems
common in some extragalactic X$_{\rm CO}$ determinations found in the
literature. In particular, we deconvolved the beam from our cloud size
estimates and we avoided unresolved clouds. Another potentially
complicating effect is cloud blending in velocity space. In poor
signal--to-noise data, emission from two or more clouds may be
confused and assigned to only one entity. Because the velocity
dispersion is squared to compute the virial mass, velocity cloud
blending can introduce large errors that overestimate virial masses.

Figure \ref{xco} shows the ratio of virial mass to molecular mass
traced by CO using the Galactic conversion factor (X$_{\rm
CO}=2\times10^{20}$ cm$^{-1}$/K km s$^{-1}$) found by using {\tt\small
CLOUDALYZE} on several extragalactic GMCs. These include observations
performed by BIMA and OVRO, as well as SEST observations of regions in
the LMC and SMC. This plot shows that resolved extragalactic GMCs have
an approximately Galactic CO--to--H$_2$ conversion factor regardless
of metallicity. The offset from M$_{virial}$/M$_{mol}=1$ is probably
due to the algorithm, which has not been scaled to produce
M$_{virial}=$M$_{mol}$ in Galactic GMCs (e.g.,
M$_{virial}$/M$_{mol}=1.5$ in the Rosette GMC).

\begin{figure}[ht]
\centering
\includegraphics[height=5.2cm]{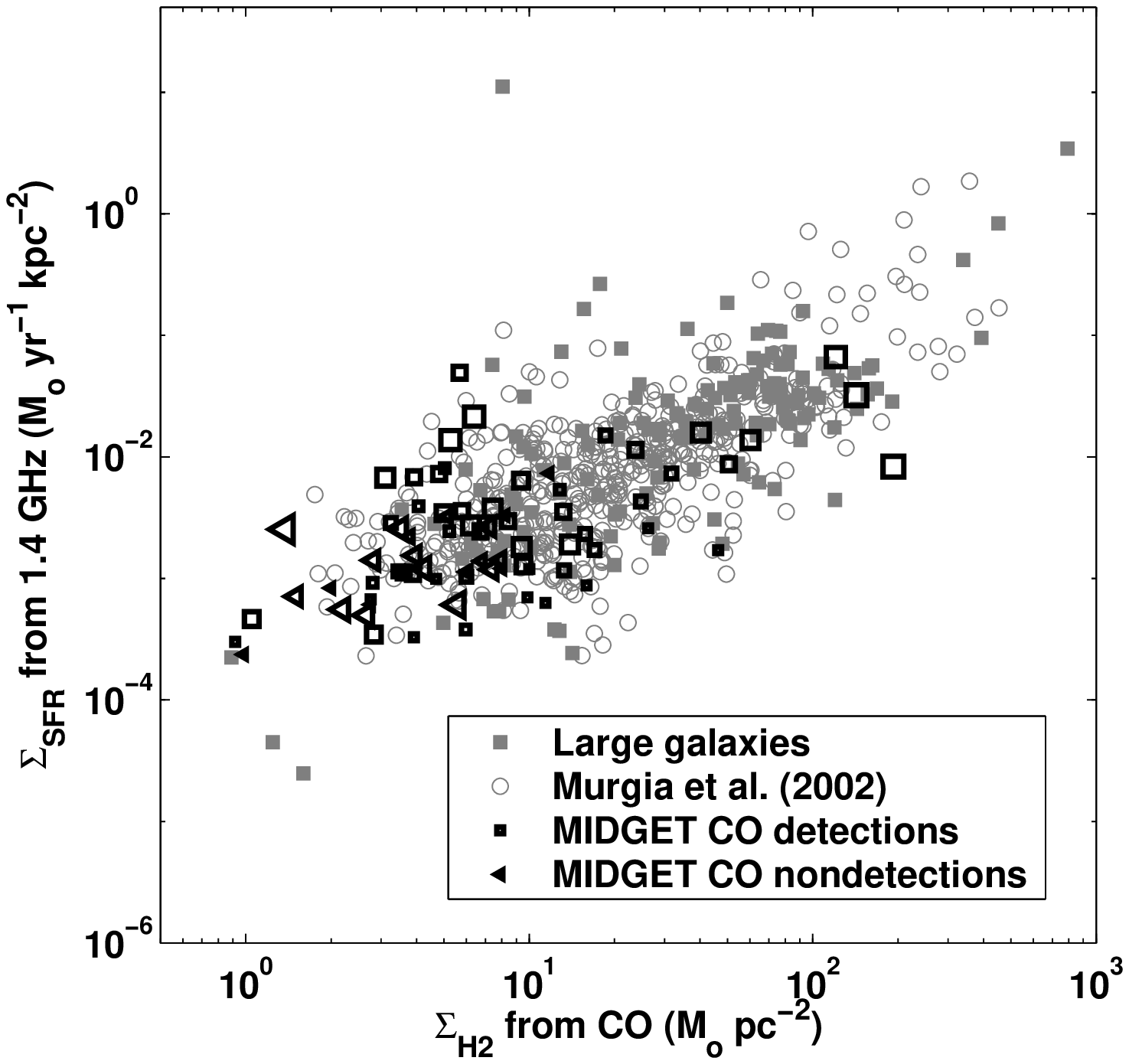}\includegraphics[height=5.2cm]{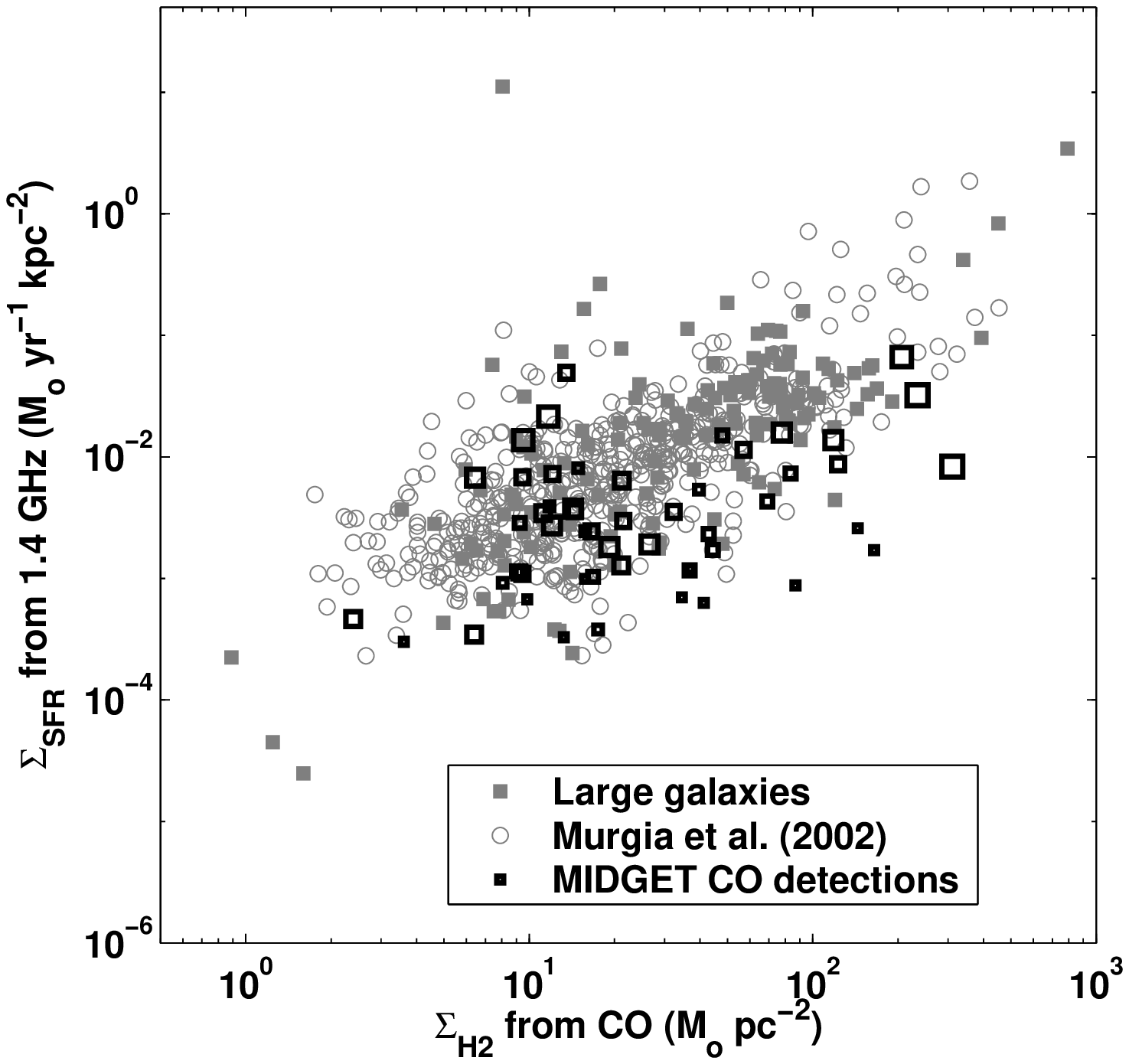}

\caption{{\bf (a)} Surface density of star formation rate versus
surface density of molecular gas, for the central pointings of
galaxies in the FCRAO catalog (gray symbols), the sample from Murgia
et al. (2002) which includes all FCRAO pointings (gray circles), and
the MIDGET galaxies (black symbols). Triangles indicate CO upper
limits. The SFR was obtained from the 1.4 GHz radio continuum NVSS
images, which have a resolution matched to that of the CO
observations. The column density of molecular gas was obtained from
the CO using the Galactic conversion factor. {\bf (b)} Same plot as
(a), but the MIDGET points now show the effects of applying a
metallicity correction to X$_{\rm CO}$ such as the one by Wilson
(1995). The symbol sizes for the MIDGET points are proportional to
metallicity.\label{rc}}
\end{figure}


Is there any evidence for a constant CO--to--H$_2$ conversion factor
apart from the virial arguments made above? Using NVSS radio continuum
images and following Murgia et al. (2002) we can readily obtain the
star formation rate (SFR) in the central regions of these galaxies
with a beam size matched to the single--dish CO measurement.  Figure
\ref{rc}a shows the surface density of SFR against that of molecular
gas. In this diagram, dwarf galaxies occupy the same locus as more
massive galaxies. If the Galactic X$_{\rm CO}$ factor used to obtain
$\Sigma_{\rm H_2}$ were systematically underestimating the amount of
H$_2$ available for star formation in dwarf galaxies then their points
would be displaced to the left of those of bigger galaxies and they
are not. In fact, attempts to correct X$_{\rm CO}$ for metallicity
effects (using the metallicities inferred from M$_{\rm B}$; Richer \&
McCall 1995) such as that in Figure \ref{rc}b clearly destroy the
agreement and appear to overestimate the surface density of molecular
gas for a given star formation rate.

It is possible that the radio continuum emission from dwarf galaxies
might be less bright for a given star formation rate --- perhaps as a
result of lower magnetic field strength in these galaxies. This effect
could collude with a metallicity-dependent X$_{\rm CO}$ factor to keep
dwarf galaxies on the same locus as the large galaxies despite the
application of an incorrect X$_{\rm CO}$ factor. We have looked into
this possibility. We find that small galaxies with V$_{rot}\sim70-110$
km s$^{-1}$ fall in the same radio continuum to far infrared
correlation as the larger galaxies: in this regard the radio continuum
seems to be as good a tracer of SFR as the far infrared for these
objects. For this alternative explanation to work, there needs to be
almost perfect cancellation between the increase in X$_{\rm CO}$ and
the reduction in the SFR measured by the radio continuum.

\section{Conclusions}

Using virial mass arguments, we showed that the X$_{\rm CO}$ factor
measured in resolved extragalactic GMCs is not strongly dependent on
metallicity. Using the correlation between radio continuum and
molecular gas, we argued that a Galactic CO--to--H$_2$ conversion
factor appears to explain the observed SFR in small galaxies as well
as in large galaxies. Thus, at least for the molecular gas relevant to
star formation processes, X$_{\rm CO}$ in these metal--poor
environments appears to be approximately Galactic regardless of
metallicity.

A few cautionary statements are in order. First, and perhaps more
importantly, this result does not mean that X$_{\rm CO}=1$ in every
conceivable environment. We are only able to obtain virial mass
estimates for very special objects, the brightest extragalactic
GMCs. This result thus may have little bearing on the CO--to--H$_2$
conversion factor over an entire galaxy, except perhaps that molecular
gas with high X$_{\rm CO}$ does not appear to be important for star
formation. Second, the physics relating the FIR and radio continuum
emission to the star formation rate are poorly understood. Therefore
it is possible that other effects come into play that cancel an
increase in X$_{\rm CO}$. This cancellation, however, needs to be
almost perfect. Finally, to properly make quantitative arguments such
as those in Figure \ref{rc}b it is desirable to have direct
metallicity measurements on a large sample of dwarf galaxies where CO
is observed. We have used a rather indirect estimator of metallicity
with a large intrinsic scatter (M$_{\rm B}$), because no other data
are available.

\printindex
\end{document}